\documentclass[11pt,a4paper]{article}
\pdfoutput=1
\usepackage{jcappub}
\usepackage[english]{babel}
\usepackage{amsfonts,amsmath,amssymb,mathrsfs}
\usepackage{graphics}
\usepackage{multirow}
\usepackage{xcolor}
\usepackage{tensor}

\newcommand{\ie}{{\it i.e.~}}
\newcommand{\eg}{{\it e.g.~}}
\newcommand{\be}{\begin{equation}}
\newcommand{\ee}{\end{equation}}
\newcommand{\ba}{\begin{eqnarray}}
\newcommand{\ea}{\end{eqnarray}}

\newcommand{\mc}{\; ,}
\newcommand{\md}{\; .}

\newcommand{\lcdm}{$\Lambda$CDM }

\newcommand{\imm}{\mathrm{i}} 
\newcommand{\rhom}{\rho_\mathrm{m}}

\newcommand{\pardern}[3]{\frac{\partial^{#1}#2}{\partial #3^{#1}}} 
\newcommand{\adj}[1]{{#1}^{\dagger}} 

\newcommand{\lagd}{\mathcal{L}}

\title{Dark matter as a Bose--Einstein Condensate: the relativistic non-minimally coupled case}
\author[a,b]{Dario Bettoni}\author[a,c]{Mattia Colombo}\author[a,b]{Stefano Liberati}
\affiliation[a]{SISSA,\\Via Bonomea 265, 34136, Trieste, Italy}
\affiliation[b]{INFN, Sezione di Trieste,\\Via Valerio 2, 34127, Trieste, Italy}
\affiliation[c]{Dep. of Physics, University of Trento, \\Via Sommarive 14, 38123 Povo (TN), Italy}
\emailAdd{bettoni@sissa.it}\emailAdd{mattia.colombo@studenti.unitn.it}\emailAdd{liberati@sissa.it}

\abstract{Bose--Einstein Condensates have been recently proposed as dark matter candidates. In order to characterize the phenomenology associated to such models, we extend previous investigations by studying the general case of a relativistic BEC on a curved background including a non-minimal coupling to curvature.  In particular, we discuss the possibility of a two phase cosmological evolution: a cold dark matter-like phase at the large scales/early times and a condensed phase inside dark matter halos. During the first phase dark matter is described by a minimally coupled weakly self-interacting scalar field, while in the second one dark matter condensates and, we shall argue, develops as a consequence the non-minimal coupling. Finally, we discuss how such non-minimal coupling could provide a new mechanism to address cold dark matter paradigm issues at galactic scales.}


\arxivnumber{1310.3753}

\begin{document}
\maketitle

\section{Introduction}

The possibility that the Cold Dark Matter (CDM) paradigm for structure formation may require modifications at small scales in order to properly account for the observed galaxy dynamics has been largely investigated in recent years. Among the most pressing issues the cosmological standard model is facing, must be mentioned the so called core-cusp problem \cite{p:deblock}, the angular momentum problem \cite{Navarro2000} and the ``too big to fail'' problem \cite{Kolchin2011,Kolchin2012,Ferrero2012}. Another source of debate is represented by the observed correlations between dark and luminous mass \cite{p:tf,p:btf1,p:btf2}, which are hard to explain in the context of the CDM paradigm. In fact, even if the introduction of baryons' feedback into pure dark matter simulations have been shown to be able to relax some of the above mentioned issues, still there is no general consensus on the effectiveness of these mechanisms \cite{p:weinberg13}. 

A nowadays popular alternative to CDM is the so called Warm Dark Matter (WDM), \ie a cosmological fluid with a small pressure, whose properties are intermediate between those of hot dark matter and cold dark matter (for example for WDM structure formation occurs bottom-up above its free-streaming scale, and top-down below its free streaming scale). Albeit this proposal has several desirable properties there seems to be a growing evidence that, in order to fit some of the extant observations, WMD has to be close to CDM to the point of being unable to solve anymore the puzzles for which it was introduced in the first place \cite{p:wdm,Viel:2013fqw}. This evidence has convinced many researchers that a sort of regime change, between small and large scales in DM dynamics, is needed in order to fit current observations.

For these reasons many alternatives to standard CDM have been proposed in recent years whose main novelty is the introduction of a new scale for DM below which the collapse into denser regions is slowed down and tamed (such a scale being different in the different mechanisms used to trigger it). All these models have to reduce to $\Lambda$CDM at scales larger than hundreds Mpc where the standard model is extremely good at explaining the cosmological evolution as testified by Cosmic Microwave Background \cite{Ade:2013zuv} and Large Scale Structure data \cite{p:lss}.

Among the competing theories stands the well known MOND paradigm \cite{p:mond} along with its relativistic extensions, notably TeVeS theories \cite{p:teves}. MOND tackles the problem by modifying the Newtonian dynamics in low acceleration regimes (the acceleration scale being given by $a_0\approx10^{-10} m\,/ s^{2}$); it has to be noted that while this model presents some problems at cluster scale and bigger, it fits remarkably well the behaviour observed in galaxies, providing also a natural explanation to the empirical relationship between the intrinsic luminosity of a spiral galaxy and its velocity width (the Tully--Fisher relation)~\cite{p:mond}. There remains however several open issues also for this framework which appears to be very effective at galactic scales but way less so for larger structures \cite{p:nomond}. 

\subsection{Non-minimally coupled DM}

A different attempt to modify the DM dynamics so to implement a scale dependence is the one advanced in \cite{p:blsf} and further explored in \cite{p:bls,p:bplb}; in these works it was shown that a non-minimal coupling of the dark matter field to gravity may allow to retrieve MOND-like phenomenology at galactic scales, while retaining the successes of the standard model of cosmology at larger scales. One may be able, in this way, to keep the successes of both $\Lambda$CDM and MOND in the regimes where they achieve the best results. 

The basic mechanism behind this proposal consist in the fact that a non-minimally coupled DM can be shown to provide a sort of effective geometry to the baryonic matter and as such it can mimic a MOND-like scenario. In particular, the coupling can be chosen in  such a way for it to be relevant only in late-time cosmology and at galactic scales, where the accelerations are low, while it is not present where accelerations are higher. In this way, for example, one could also explain the apparent need for some amount of dark matter in clusters in MOND; indeed in the centre of clusters, where accelerations are higher, the behaviour is still as for CDM, while in the outskirts accelerations are low enough to permit the dynamics to be MONDian. 

In \cite{p:bls} the model was further expanded considering, instead than just a field, a fluid non-minimally coupled to gravity. In this way it was possible to derive a Poisson equation in the Newtonian limit where the gravitational potential was found to be not anymore dependent just on the density, but also on gradients of the density. This allows to modify the dynamics in such a way that it lets the formation of structures to be enhanced or suppressed as compared to the standard scenario, thus in principle permitting to suitably fit the observed structure formation.  Of course, as said, for such a model to be viable it was required the non minimal coupling might to develop in a dynamical way. Noticeably, among the possible mechanisms, it was proposed the Bose-Einstein condensation of the dark matter fluid \cite{p:bls}.\footnote{See \cite{Folkerts:2013tua} for an application of NMC to the case of axion DM.}

\subsection{DM as a BEC}

The bold proposal of DM as a Bose--Einstein condensate (BEC),  has recently gained some attention as a possible alternative to the CDM framework within which conciliate many of the thorny issues previously described. The phenomenon of boson condensation was theoretically predicted almost ninety years ago \cite{p:bose,p:einstein1,p:einstein2}, but it became part of experimental physics only seventy years later \cite{p:cornell} and since then its properties have been largely investigated in laboratory \cite{p:kett}. A BEC consists in a large number of boson particles attaining a macroscopic occupation number of the ground state when the system is below some critical temperature. When this happens the quantum particle nature of the ground state is substituted by a collective one in terms of a classical scalar field, the condensate wave function, with the quantum nature confined to small phononic perturbations above the ground state. A remarkable fact is that, during the condensate phase, the BEC admits a hydrodynamical description in terms of the particle density and velocity potential, and its equations closely resemble the continuity and Euler ones. This makes the extension to the astrophysical/cosmological context straightforward even though some differences must be considered. 

While the first appearance of BEC in the astrophysical context dates back to 1983 \cite{p:bald}, the application to dark matter has been proposed only some years later \cite{p:sin,p:bohmerharko,p:fmt,p:harkoa,p:harkob,p:chaa,p:chab,p:rds}. In these works BEC has been described mainly via a non-relativistic, time-dependent Gross--Pitaevskii equation characterised by an external potential (the trap) mimicking the gravitational potential associated with a galaxy.  Such gravitational potential is generally determined by coupling the Poisson equation to the Gross--Pitaevskii one and assigning a polytropic equation of state to the condensate. 

Some advantages of this scenario has been already pointed out in the aforementioned references. First of all, in \cite{p:harkoa} it is shown how the Bose--Einstein condensate dark matter model may be able to solve the core/cusp problem while in \cite{p:mul} a solution to the overabundance of structures is proposed. Secondly, in \cite{p:bohmerharko} it has been shown, for a sample of Low Surface Brightness galaxies and dwarf galaxies, that the rotation curves obtained form BEC DM are in agreement with the observed ones, even though in \cite{Guzman:2013coa,Guzman:2013rua} it has been pointed out how the solution found in \cite{p:bohmerharko} may not be stable. Furthermore, the bending angle of light in gravitational lensing is found to be much larger than the value predicted by standard dark matter models. This could be therefore a discriminating feature to test the validity of the model and indicates that the inclusion of BEC DM in a more formalized setup is required. Further analysis of this framework phenomenology can be found in \cite{p:fmt,p:harkoa,p:harkob,p:chaa,p:chab,p:rds,p:mul} while an extensive review of such models is found in \cite{p:srm}. 

\subsection{Non-minimally coupled BEC as DM}

In this paper we shall try to further explore the proposal put forward in \cite{p:bls} by explicitly considering the possible role of Bose--Einstein condensation in providing a dynamical mechanism for the development of a non-minimal coupling. To do so we shall necessarily have to go beyond the so far accomplished BEC DM analyses~\cite{p:sin,p:bohmerharko,p:fmt,p:harkoa,p:harkob,p:chaa,p:chab,p:rds} as these studied mainly the dynamics of a non-relativistic condensate. This is not quite the right ansatz to fully couple a BEC to gravity and most of all cannot entail the possibility of non-minimal couplings.\footnote{Actually there is an example of a model trying to treat the condensate as relativistic \cite{p:fm}, but also in this case the relativistic treatment is limited to analyzing the complex scalar field in a field theoretical guise and the derived relativistic Gross--Pitaevskii equation is not consistently used in a general relativistic framework.} Furthermore, most of these models end up requiring some sort of ultralight boson, with mass of order $10^{-22}$ eV, to provide the required pressure support to produce suitably cored profiles and to suppress small scale power spectrum \cite{Hu:2000ke}.\footnote{See however \cite{Woo:2008nn} for a counterexample.} However, with such a small mass one might wonder if considering the condensate as non-relativistic is a justified assumption.

We shall then consider a fully relativistic Bose--Einstein condensate coupled to gravity, including a non-minimal coupling, so to explore the viability of the mechanisms  proposed in \cite{p:blsf,p:bls} and most of all explore the dynamics implied by such a complete framework.
The paper is organized in the following way: in section \ref{FlatRBEC} we give a brief recap of the relativistic BEC in flat space, then in section \ref{NMCRBEC} we construct the model for non-minimally coupled BEC in a general relativistic framework, deriving the equations of motion for the condensate and the modified Einstein equations and investigating the Newtonian limit, in particular the resulting modified Poisson equation while in section  \ref{pheno} we discuss the phenomenology expected. Finally in section \ref{conclusions} we shall draw our conclusions. 

\section{Elements of relativistic BEC in flat space}
\label{FlatRBEC}

In this section we will briefly introduce the formalism for relativistic BEC referring to \cite{p:fagnocchi} and references therein for details.

The Lagrangian density for an interacting relativistic scalar Bose field $\hat{\phi}(\vec{x},t)$ is 
\begin{equation}\label{eq:lagdfs}
\lagd=\frac{1}{c^2}\pardern{}{\adj{\hat{\phi}}}{t}\pardern{}{\hat{\phi}}{t}-\vec{\nabla}\adj{\hat{\phi}}\cdot\vec{\nabla}\hat{\phi}-\left( \frac{m^2 c^2}{\hbar^2}+V(t,\vec{x}) \right)\hat\rho-U(\hat\rho,\lambda_i)
\end{equation}
being $V(\vec{x},t)$ some external potential and U a self-interaction term in the form
\begin{equation}\label{}
U(\hat\rho,\lambda_i)=\frac{\lambda_2}{2}\hat{\rho}^2+\frac{\lambda_3}{6}\hat{\rho}^3+\dots
\end{equation}
where $\hat{\rho}=\adj{\hat{\phi}}\hat{\phi}$ is the density operator and $\lambda_i$ are the coupling constant which may also depend on spacetime coordinates. The first term corresponds to the usual two particle interaction $\lambda\phi^4$ while the other terms in the series corresponds to many body interactions.

The Lagrangian density \eqref{eq:lagdfs} is invariant under a global $U(1)$ transformations, whose associated conserved current is
\be
j^\mu=\frac{\imm}{2}\left(\hat\phi^\dagger\partial^\mu \hat\phi-\hat\phi\partial^\mu\hat\phi^\dagger\right)\,.
\ee
to which corresponds the conserved ensamble charge $N-\bar N$, where $N$ $(\bar N)$ is the number of bosons (anti-bosons).  
For a non interacting BEC the conserved charge can be written in terms of the average number of bosons $n_k$ and anti boson $\bar n_k$ in an energy state $E_k$ \cite{Grether:2007ur}
\be
N-\bar N=\sum_k n_k-\bar n_k\,,
\ee
 where 
 \be 
 n_k(\mu,\beta)=1/\{\text{exp}[\beta(|E_k|-\mu)]-1\}\,, \quad \bar n_k(\mu,\beta)=1/\{\text{exp}[\beta(|E_k|+\mu)]-1\}\,,
 \ee
where $\mu$ is the chemical potential, $E^2_k=\hbar^2c^2k^2+m^2c^4$ is the energy of the state $k$ and $\beta=k_B T$, being $T$ the temperature and $k_B$ the Boltzmann constant.
This allows to find the relation between the critical temperature and the conserved charge density $n\equiv (N-\bar N)/V$, where $V$ is the volume of the system,
\be
n= C\int_0^\infty dk k^2 \frac{\sinh(\beta_c mc^2)}{\cosh(\beta_c |E_k|)-\cosh(\beta_cmc^2)}\,,
\ee
where we have used the fact that at the critical temperature the chemical potential is $\mu=mc^2$ \cite{Grether:2007ur} and where C is a constant given by
\be
C= \frac{1}{4\pi^{3/2}\Gamma(3/2)}\,.
\ee
From this expression can be derived the ultrarelativistic and non-relativistic relations between the critical temperature and the density of particles in a given state, namely
\ba\label{tcrit}
k_BT_c^{NR}&=&\frac{2\pi \hbar^2}{m}\left[\frac{n}{\zeta(3/2)}\right]^{2/3}\,,\\
k_BT_c^{UR}&=&\left[\frac{\hbar^3c\Gamma(3/2)(2\pi)^3}{4m\pi^{3/2}\Gamma(3)\zeta(2)}\right]^{1/2}n^{1/2}\,,
\ea
where $\zeta$ is the Riemann zeta function and $m$ is the mass of the boson.

When $T\gg T_c$ the behavior of the scalar field is that of a standard self-interacting quantum field. However, when $T\ll T_c$ the scalar field undergoes a phase transition and almost all particles condense and end up occupying the ground state of the system. During this phase the condensate can be described as
\be
\hat\phi =\phi(1+\hat\varphi)
\ee
where $\phi$ is the condensate wave function, a classical field which describes the collective behavior of the ground state, while $\hat \varphi$ represents quantum excitations. With this decomposition and neglecting exited states, the non-linear Klein--Gordon equation becomes
\begin{equation}\label{ClassicalKG}
\Box_{\mathrm{f}}\phi+\left( \frac{m^2 c^2}{\hbar^2}+V(t,\vec{x}) \right)\phi+U'\phi=0\mc
\end{equation}
where prime indicates derivatives with respect to $\rho$ and where we have introduced the Minkoski metric $\eta_{\mu\nu}=\text{diag}[-1,1,1,1]$,  the subscript $\rm{f}$ indicates we are working in flat spacetime, so that
\be
\Box_{\mathrm{f}} \phi=\eta^{\mu\nu}\partial_\mu\partial_\nu\phi=-\frac{1}{c^2}\pardern{2}{\phi}{t}+\nabla^2\phi\,.
\ee
 We stress that equation \eqref{ClassicalKG} exactly reduces to the Gross--Pitaevskii one in the non-relativistic limit, as shown below.

In the condensate phase the BEC wave function $\phi$ admits a useful description in terms of the Madelung representation, in which the classical complex field is split into an amplitude $\rho$ and a phase $\theta$
\begin{equation}\label{eq:mad}
\phi=\frac{\hbar}{m}\sqrt{\frac{\rhom}{2}}\,e^{\imm \theta}\,,
\end{equation}
where we have defined 
\be
\rho_\text{m}\equiv \frac{2m^2}{\hbar^2}\rho
\ee
which has the dimensions of a mass density.
This redefinition can always be used, being $\rho$ and $\theta$ the probability density and the phase of the field, but in the case of a BEC they assume the physical meaning of the measured density and of velocity potential, in the non-relativistic limit.

Substituting these variables in the Klein--Gordon equation and in the current conservation and defining the 4-vector $u^\mu=\hbar/m\nabla^\mu\theta$, we obtain the fluid equations 
\begin{eqnarray}\label{eq:flatfleom}
\nabla^\mu(\rhom u^\mu)&=&0\\
-u^\mu u_\mu&=&c^2+\left(2\,U'(\rhom)-\frac{\hbar^2}{m^2}\frac{\Box_\mathrm{f} \sqrt{\rhom}}{\sqrt{\rhom}}\right)\,,
\end{eqnarray}
where from now on the prime indicates derivatives with respect to $\rhom$. The first equation represents the conservation of the current $j^\mu=\rhom \hbar u^\mu/m$ and closely resembles the standard continuity equation for fluids. However, the second one is not the Euler equation but rather a constraint equation for the norm of the 4-vector $u^\mu$, whose zero component $u^0$ can be shown to be related to the chemical potential \cite{p:fagnocchi}. In fact, $u^\mu$ is not the standard 4-velocity, as is clear from the fact that it is not normalized to unity as for perfect fluids. Indeed, $u^\mu$ is related to the fluid 4-velocity $v^\mu$ by
\be
v^\mu=\frac{u^\mu}{\sqrt{u^\mu u_\mu}}
\ee
as is discussed in appendix \ref{Flrepr}, where also relations between hydrodynamical and BEC variables are computed.
Despite this, it is possible to write the second equation in \eqref{eq:flatfleom} in the form of an evolution equation for $u^\mu$, by taking the derivative with respect to $\partial_\nu$
\be
-u^\mu\partial_\mu u_\nu=\partial_\nu U' -\partial_\nu\left(\frac{\hbar^2}{2m^2}\frac{\Box_\mathrm{f}\sqrt{\rhom}}{\sqrt{\rhom}}\right)\,,
\ee
where we have used the property $\partial_\nu u_\mu=\partial_\mu u_\nu$ which can be easily derived from the definition of $u^\mu$. We will see in the following how this equation coincides with the Euler equation when quantum properties of the BEC can be neglected (Thomas--Fermi limit). To better appreciate this, it is useful to rewrite equation \eqref{eq:flatfleom} as
\be 
-u^\mu u_\mu=c^2-U'\left(1-2\xi^2\frac{\Box_\mathrm{f} \sqrt{\rhom}}{\sqrt{\rhom}}\right)\,,
\ee
where we have introduced, in analogy with the non-relativistic case, the healing length of the condensate $\xi^2=\hbar^2/(2m^2U')$ which characterises the typical scale of time/space variations of the condensate amplitude. Whenever these happen on scales quite larger than $\xi$ we can neglect the quantum pressure term and the Thomas--Fermi limit applies.\footnote{For a detailed description of this issue in the non-relativistic limit see \cite{b:ps}.}\\
As a concluding remark we recall that the system \eqref{eq:flatfleom} is completely equivalent to the non-linear Klein--Gordon equation \eqref{ClassicalKG}.

\section{Non-minimally coupled relativistic BEC}
\label{NMCRBEC}

We now proceed to the generalization of the relativistic BEC theory in the case of curved spacetimes. The standard procedure employs the minimal coupling principle, just substituting the Minkowski metric with a spacetime dependent one and the derivatives with covariant ones. However, underlying this procedure there is the assumption that matter does not couple directly to second derivatives of the metric thus allowing to neglect them in small enough regions of spacetime. But this may not be the case if the physics of the matter component is such that it directly probes curvature, thus being able to distinguish at a fundamental level if spacetime is curved or not.

We have seen that one of the characteristics of a BEC is the existence of a typical length scale, the healing length  $\xi$, that regulates the minimal size at which the classical, collective, fluid description is valid through the quantum pressure term in equation \eqref{velNRhealing}. As such, it forbids shocks and in general sharp changes in the density distribution on scales smaller than $\xi$. This physical characterisation of the BEC can be easily extended to relativistic BECs (see e.g.~\cite{p:fagnocchi}). The question we want to address now is what happens when one generalizes from flat to curved spacetimes the dynamics for a fluid endowed with a macroscopic coherence length.  In particular, for a BEC when moving to curved spacetimes the problem changes from a single scale (the healing length) to a two scales problem (healing and curvature). Hence, it should be expected the presence of extra interactions between BEC and geometry besides the ones provided by the volume element when the two scales are of the same order of magnitude.

We argue here that, when in the condensed phase, the scalar field must couple to curvature terms with a characteristic length scale provided by the healing length. There are not many ways that this can be implemented with a scalar field requiring second order field equations for both the metric and the scalar field. Indeed only two couplings are possible~\footnote{Another possibility is to consider a conformal coupling of the type $\rho R$, but in this case no intrinsic length scale appears and it is not clear why the condensation should lead to such coupling if previously absent.} 
\be
XR\quad\text{ and }\quad G_{\mu\nu}\nabla^\mu\phi\nabla^\nu\phi\,,
\ee
where $\phi$ is the BEC field and $X=\nabla_\mu\phi\nabla^\mu\phi/2$ is its kinetic term.
However, these two terms are equivalent modulo a surface term \cite{p:bl} and hence we will only deal with the one proportional to the Einstein tensor $G_{\mu\nu}$. This, in order to have the correct dimensions, must be multiplied by a constant with the dimensions of a length. 

This is a crucial point because we argue that such non-minimal coupling should always be considered when the scalar field is in the condensate phase and its characteristic length is of the same order as the curvature one. On the contrary, in the non condensed phase, such term cannot be generated dynamically given that before condensation the healing length does not exist and hence no natural scale can be introduced.\footnote{Notice that even if from a field theoretical point of view nothing forbids the presence of such coupling between scalar field and the Einstein tensor, with the limitations imposed by the Horndeski theorem, to be present also in the non condensed phase, such coupling would be present at all scales and all times (see \eg~ \cite{Folkerts:2013tua}).} Also, this issue points out a fundamental difference with the other cosmological fluids: for these there is no cosmologically relevant length scale\footnote{Actually all fluids have a characteristic length, the mean free path. For cosmological fluids this is however so small, compared with astrophysical sizes, that can be safely neglected. This is obviously not the case of CDM, which being pressureless has in principle an infinite mean free path, and hence does not admit a proper fluid limit. Of course if one assumes DM to be very weakly interacting then it would admit a finite mean free path. However, in this case it is not clear if this would be short enough so to be negligible on the typical scales on which the fluid approximation is used.} and hence their fluid description can be made in the flat spacetime limit.

For this reason we will investigate a model in which, besides the standard generalizations to curved spacetimes, we introduce a non-minimal coupling between the scalar field derivatives and the Einstein tensor in the form
\be
L^2 G_{\mu\nu}\nabla^\mu\adj{\hat \phi}\nabla^\nu\hat \phi\,,
\ee
where $L$ is a some length scale characteristic of the system that we shall expect to be of the order of the BEC healing length $\xi$.\footnote{In principle we could have considered a more general coupling by including a generic function $f(\adj{\phi}\phi)$ in front of the Einstein coupling. However, for the economy of this work it is enough to consider the above term, even if the more general coupling has to be considered when comparing the predicted phenomenology with real data.}

\subsection{Action and equation of motions}

The non-minimally coupled relativistic BEC just introduced is described by the following action
\be\label{eq:act}
S=\int d^{4}x\sqrt{-g}\left[\frac{c^3}{16\pi G_N}R+\frac{1}{c}\mathcal{L}_{\phi}\right]+\epsilon\int d^{4}x\sqrt{-g}\left[\frac{L^{2}}{c}G_{\mu\nu}\nabla^{\mu}\hat \phi^{\dagger}\nabla^{\nu}\hat \phi\right]\mc
\ee
where $G_{\mu\nu}$ is the Einstein tensor, $G_N$ is the gravitational constant and $\epsilon$ is a unitary dimensionless constant which can take up values $\epsilon=\pm1$, being the sign of the non-minimal coupling not determined a priori; this constant is also useful as a sort of bookeeping parameter to easily keep track of the effects of the NMC. The Lagrangian density for the Bose field is the generalization to curved spacetimes of \eqref{eq:lagdfs}
\be
\lagd_{\phi}=-g_{\mu\nu}\nabla^{\mu}\hat \phi^{\dagger}\nabla^{\nu}\hat \phi-\frac{m^{2}c^{2}}{\hbar^{2}}\hat \rho -U(\hat \rho)\,.
\ee
Notice that, differently from the flat space case, we are not considering any external potential because in our case there are no external fields other than the gravitational one, which is already taken into account by the coupling of the field to the metric and its derivatives.

Taking the variation of the action \eqref{eq:act} with respect to the scalar field we obtain the modified Klein--Gordon equation for the condensate wave function
\be\label{eq:eomf}
\left(g_{\mu\nu}-\epsilon L^{2}G_{\mu\nu}\right)\nabla^{\mu}\nabla^{\nu}\hat\phi-\left(\frac{m^{2}c^{2}}{\hbar^{2}}+U'(\hat\rho)\right)\hat\phi=0\,.
\ee
Here noticeably, the structure of the equation is  the same we would obtain with a minimally coupled field with the exception of the derivative structure which is changed by the presence of the NMC. In particular, we can define an effective derivative operator
\be
\bar \Box = \left(g_{\mu\nu}-\epsilon L^{2}G_{\mu\nu}\right)\nabla^{\mu}\nabla^{\nu}\,,
\ee
which contains all the effects of the non-minimal coupling. As a consequence we have that, compared with a minimally coupled BEC, the acceleration is reduced or enhanced by the NMC for a fixed source. However, one must be careful about possible zeros of the effective derivative operator. In particular, at the leading order in the $\epsilon$ parameter, we have
\be
\bar \Box = \left(g_{\mu\nu}-\frac{8\pi G_N}{c^4}\epsilon L^{2}T^\text{matter}_{\mu\nu}+ \mathcal{O}(\epsilon^2)\right)\nabla^{\mu}\nabla^{\nu}\,,
\ee
which shows how the change in the modulus of the derivative operator depends on the matter content and that for very high density regions the operator may become singular. Anyway, as already discussed in a different context \cite{Zumalacarregui:2012us}, only a full dynamical analysis can state under which conditions this can happen (and eventually be used to fix the sign of the NMC, $\epsilon$). Moreover, the weak field limit (which will be mostly relevant for the confrontation with observations) is in the end free of such ambiguities.

The variation of $\mathcal{L}_{\phi}$ with respect to the metric gives the stress-energy tensor (SET) for the minimally coupled field
\be
T_{\mu\nu}^{\phi}=\nabla_{\mu}\phi^{\dagger}\nabla_{\nu}\phi+\nabla_{\mu}\phi\nabla_{\nu}\phi^{\dagger}-g_{\mu\nu}\left(g^{\alpha\beta}\nabla_{\alpha}\phi^{\dagger}\nabla_{\beta}\phi+\frac{m^{2}c^{2}}{\hbar^{2}}\rho+U(\rho)\right)\,,
\ee
while the same variation of the non-minimal coupling part gives 
\be\label{eq:setnmc}
\begin{split}
T^{\mu\nu}_{\mathrm{NMC}}&=L^2\left[g^{\mu\nu}G_{\alpha\beta}\nabla^{\alpha}\phi^{\dagger}\nabla^{\beta}\phi+R^{\mu\nu}g_{\alpha\beta}\nabla^{\alpha}\phi^{\dagger}\nabla^{\beta}\phi-R\nabla^{\mu}\phi^{\dagger}\nabla^{\nu}\phi\right.\\
&\phantom{=}+\left.g^{\sigma\mu}\nabla_{\alpha}\nabla_{\sigma}(\phi^{\dagger\alpha}\phi^{\nu}+\phi^{\dagger\nu}\phi^{\alpha})-\square(\phi^{\dagger\mu}\phi^{\nu})-g^{\mu\nu}\nabla_{\alpha}\nabla_{\beta}(\phi^{\dagger\alpha}\phi^{\beta})\right.\\
&\phantom{=}\left.+(g^{\mu\nu}\square-\nabla^{\mu}\nabla^{\nu})(\phi^{\dagger\alpha}\phi_{\alpha})\right]
\end{split}
\ee
with $\phi^\alpha\equiv\nabla^\alpha\phi$. This contribution to the SET contains curvature terms, as expected, and field derivatives up to third order. However, it has to be noted how these higher derivatives can be eliminated in favor of curvature terms so that both the Einstein field equations and the field equations derived from the covariant conservation of the total SET are at most second order. This is a natural consequence of the fact that the action \eqref{eq:act} is a subcase of the Horndeski one~\cite{p:bl}.

Finally, the Einstein field equations are
\be\label{eq:ein}
\tensor{G}{_{\mu\nu}}=\frac{8\pi G}{c^{4}}\left[T_{\mu\nu}^{\phi}+\epsilon T_{\mu\nu}^{\mathrm{NMC}}\right]\md
\ee
We stress that, despite the appearances, these equations are not the standard one as in the second term on the RHS are hidden curvature quantities.

\subsection{Fluid representation}

In section \ref{FlatRBEC} we have pointed out how, when $T\ll T_c$, the ground state of the condensate can be described by a classical complex field $\phi$ that can be rexpressed with the fluid variables $\rho$ and $\theta$. 
When expressed in terms of these we obtain for the minimally coupled SET of the field
\begin{equation}\label{eq:setmad}
T^{\phi}_{\mu\nu}=\rho_\text{m} u_\mu u_\nu-\tensor{g}{_{\mu\nu}}\left(\rho_\text{m}\frac{c^2+u^2}{2}+\frac{\hbar^2}{2m^2}\nabla^\alpha\sqrt{\rho_\text{m}}\,\nabla_\alpha\sqrt{\rho_\text{m}}+U(\rho_\text{m})\right)+\frac{\hbar^2}{m^2}\nabla_\mu\sqrt{\rho_\text{m}}\,\nabla_\nu\sqrt{\rho_\text{m}}
\end{equation}
and for the stress-energy tensor of the non-minimal coupling term
\begin{equation}\label{eq:setnmcmad}\begin{split}
T^{NMC}_{\mu\nu}&=\frac{L^2}{2}\left[g_{\mu\nu}G_{\alpha\beta}\left(\rhom u^\alpha u^\beta+\frac{\hbar^2}{m^2}\nabla^\alpha\sqrt{\rhom}\,\nabla^\beta\sqrt{\rhom}\right)\right.\\
&\phantom{=}+\left.+R_{\mu\nu}\left(\rhom u^2+\frac{\hbar^2}{m^2}\nabla^\alpha\sqrt{\rhom}\,\nabla_\alpha\sqrt{\rhom}\right)\right.\\
&\phantom{=}+\left.R\left(\frac{\hbar^2}{m^2}\nabla_\mu\sqrt{\rhom}\,\nabla_\nu\sqrt{\rhom}+\rhom u_\mu u_\nu\right)\right]\\
&\phantom{=}+2L^2\nabla_\mu\nabla^\alpha\left(\frac{\hbar^2}{m^2}\nabla_{\alpha}\sqrt{\rhom}\,\nabla_{\nu}\sqrt{\rhom}+\rhom u_\alpha u_\nu\right)\\
&\phantom{=}-L^2\square\left(\frac{\hbar^2}{m^2}\nabla_{\mu}\sqrt{\rhom}\,\nabla_{\nu}\sqrt{\rhom}+\rhom u_\mu u_\nu\right)\\
&\phantom{=}-L^2\tensor{g}{_{\mu\nu}}\nabla^\alpha\nabla^\beta\left(\frac{\hbar^2}{m^2}\nabla_{\alpha}\sqrt{\rhom}\,\nabla_{\beta}\sqrt{\rhom}+\rhom u_\alpha u_\beta\right)\\ 
&\phantom{=}+L^2(\tensor{g}{_{\mu\nu}}\square-\nabla_\mu\nabla_\nu)\left(\frac{\hbar^2}{m^2}\nabla^{\alpha}\sqrt{\rhom}\,\nabla_{\alpha}\sqrt{\rhom}+\rhom u^\alpha u_\alpha\right)\,,
\end{split}\end{equation}
where we have introduced the squared norm of the 4-vector $u^\alpha$, $u^2\equiv u^\alpha u_\alpha$.
 This notation is particularly useful because shows how, when we neglect terms proportional to $\hbar^2$, \ie quantum terms, and NMC, we recover the SET for a standard perfect fluid as discussed in appendix \ref{Flrepr}.

Substituting \eqref{eq:mad} inside the equation of motion \eqref{eq:eomf} we obtain the equation of motion for the condensate in the fluid limit\footnote{These equations can also be derived by writing the conserved current $j^\mu$ and the SET in terms of $\rho$ and $\theta$ and use their covariant conservations.}
\begin{eqnarray}\label{eq:fleom}
\phantom{-}\left(\tensor{g}{_{\mu\nu}}-\epsilon L^2\tensor{G}{_{\mu\nu}}\right)\nabla^\nu(\rho_\text{m} u^\mu)&=&0\,,\\
-\left( \tensor{g}{_{\mu\nu}}-\epsilon L^2\tensor{G}{_{\mu\nu}} \right)u^\mu u^\nu&=&c^2+2U'(\rhom)-\frac{\hbar^2}{m^2}\left( \tensor{g}{_{\mu\nu}}-\epsilon L^2\tensor{G}{_{\mu\nu}} \right)\frac{\nabla^\mu\nabla^\nu \sqrt{\rho_\text{m}}}{\sqrt{\rho_\text{m}}}\,.
\end{eqnarray}
As we have already discussed, the NMC changes the derivative structure of the equations with respect to  the flat case \eqref{eq:flatfleom} given that indices contractions are made with respect to the effective metric $\bar g_{\mu\nu}=\tensor{g}{_{\mu\nu}}-\epsilon L^2\tensor{G}{_{\mu\nu}} $ which is built with the Einstein tensor. This in turns is related to the BEC SET through the Einstein equations, meaning that the effect of the NMC is to introduce a new dependence on the local BEC fluid variables in the equations of motion. Interestingly, the non-minimal coupling affects the relativistic quantum pressure term potentially enhancing its effects as compared to the flat space case.

We will not discuss these equations in their full generality here, leaving it for further studies, even if we will anticipate some of the expected phenomenology in section \ref{pheno}.
Instead, in the next section, we will see how the modification introduced by the NMC to the Einstein--Klein--Gordon system affect the dynamics in the non-relativistic regime.

\subsection{Newtonian limit}

The Newtonian regime is obtained by taking the weak field limit of the gravitational interaction. This is a meaningful approximation for low density and small velocity physical systems and is particularly suited for the investigation of gravitational dynamics at galactic scales. In order to do so we need to properly expand the metric $\tensor{g}{_{\mu\nu}}\to\tensor{\eta}{_{\mu\nu}}+\tensor{h}{_{\mu\nu}}$ where $\tensor{\eta}{_{\mu\nu}}$ is the Minkowski metric and $\tensor{h}{_{\mu\nu}}$ is a small perturbation. It is then convenient to define
\begin{equation}\label{eq:htransf}
\tensor{\bar{h}}{_{\mu\nu}}=\tensor{h}{_{\mu\nu}}-\frac{1}{2}\tensor{\eta}{_{\mu\nu}}h \quad\rightarrow\quad \tensor{h}{_{\mu\nu}}=\tensor{\bar{h}}{_{\mu\nu}}-\frac{1}{2}\tensor{\eta}{_{\mu\nu}}\bar{h}
\end{equation}
and using the transverse gauge $\partial^\mu\tensor{\bar{h}}{_{\mu\nu}}=0$, we have at first order
\begin{equation}\label{eq:einstein}
G^{(1)}_{\mu\nu}\equiv-\frac{1}{2}\square\tensor{\bar{h}}{_{\mu\nu}}=\frac{8\pi G_N}{c^4}\tensor{T}{_{\mu\nu}} \quad\rightarrow\quad \square\tensor{\bar{h}}{_{\mu\nu}}=-\frac{16\pi G_N}{c^4}\left[T_{\mu\nu}^{\phi}+\epsilon T_{\mu\nu}^{NMC}\right]\md
\end{equation}
In order for weak gravity to be a consistent description we need to consider low densities and slow motion so that time derivatives can be neglected when compared to spatial ones.
With these considerations, the SET has to be taken at zeroth order in $h_{\mu\nu}$  thus making the curvature terms in \eqref{eq:setnmcmad} vanish. One then gets
\begin{equation}\label{eq:setmadflat}
\tensor{T}{_{\mu\nu}}=\rhom u_\mu u_\nu-\tensor{\eta}{_{\mu\nu}}\left(\rhom\frac{c^2+u^2}{2}+\frac{\hbar^2}{2m^2}\partial^\alpha\sqrt{\rhom}\,\partial_\alpha\sqrt{\rhom}+U(\rhom)\right)+\frac{\hbar^2}{m^2}\partial_\mu\sqrt{\rhom}\,\partial_\nu\sqrt{\rhom}\mc
\end{equation}
\begin{equation}\label{eq:setnmcmadflat}\begin{split}
T^{NMC}_{\mu\nu}&=L^2\partial_\mu\partial^\alpha\left(\frac{\hbar^2}{m^2}\partial_{\alpha}\sqrt{\rhom}\,\partial_{\nu}\sqrt{\rhom}+\rhom u_\alpha u_\nu\right)\\
&\phantom{=}-\frac{L^2}{2}\square\left(\frac{\hbar^2}{m^2}\partial_{\mu}\sqrt{\rhom}\,\partial_{\nu}\sqrt{\rhom}+\rhom u_\mu u_\nu\right)\\
&\phantom{=}-\frac{L^2}{2}\tensor{\eta}{_{\mu\nu}}\partial^\alpha\partial^\beta\left(\frac{\hbar^2}{m^2}\partial_{\alpha}\sqrt{\rhom}\,\partial_{\beta}\sqrt{\rhom}+\rhom u_\alpha u_\beta\right)\\
&\phantom{=}+\frac{L^2}{2}(\tensor{\eta}{_{\mu\nu}}\square-\partial_\mu\partial_\nu)\left(\frac{\hbar^2}{m^2}\partial^{\alpha}\sqrt{\rhom}\,\partial_{\alpha}\sqrt{\rhom}+\rhom u^2\right)\md
\end{split}\end{equation}

To compute the Newtonian limit we consider the form of \eqref{eq:einstein} and recall that from \eqref{eq:flatfleom} in the non-relativistic limit the fluid variables behave like $u^0\to c$ and $\vec{u}\to\vec{v}$ (3-velocity of the fluid) which dramatically reduces the number of terms that are relevant in the weak field limit. 
In particular, the Poisson equation can be derived from the $(00)$ component of the linearized Einstein equations. Identifying $h_{00}\equiv-2\Phi_N/c^2$, we obtain 
\begin{equation}\label{eq:poisson}
\nabla^2\Phi_N=4\pi G_N \left(\rhom-\epsilon L^2{\nabla^2\rhom}\right)\md
\end{equation}
This result is analogous to the one obtained in \cite{p:bls} for standard NMC perfect fluids. However, in that case, the particular coupling chosen here would lead to no corrections in the weak field limit in contrast with what obtained here thus signalling a difference in the behavior of standard fluids with respect to BECs. This is evident if we compare the shape of the NMC in the two cases. For a perfect fluid we have
\be\label{eq:perfectfluid}
f(\rho)G_{\mu\nu}v^\mu v^\nu
\ee
where $v^\mu$ is the fluid four velocity. It can be shown that such coupling will lead to no contributions to the Poisson equation in the non relativistic limit (see \cite{p:bls}). For a BEC the coupling reads
\be
G_{\mu\nu}\nabla^\mu\phi\nabla^\nu\phi^\dagger = \frac{\rho_m}{2}G_{\mu\nu}u^\mu u^\nu + \frac{\hbar^2}{2m^2}G_{\mu\nu}\nabla^\mu\sqrt\rho_m\nabla^\nu\sqrt\rho_m
\ee
where besides the standard fluid coupling there is a new contribution related to the quantum nature of the BEC which is absent in the case of as perfect fluid \eqref{eq:perfectfluid}.
Finally, notice that, even if no explicit quantum term is present in the weak field limit, its presence it is nonetheless hidden in $L^2$ as we will discuss in section \ref{pheno}. 

The $(ij)$ components of the Einstein equations are
\be
\nabla^2 h_{ij} = -\frac{8\pi G_N}{c^2}\left[\rhom\left(1 +\epsilon L^2\frac{\nabla^2\rhom}{\rhom}\right)\delta_{ij} + \epsilon L^2\partial_i \partial_j\rhom\right]\,.
\ee 
This potential is composed by a diagonal part plus an anisotropic part and it is a new potential besides the Newtonian one thus making clear that the NMC excites new gravitational degrees of freedom in the weak field limit.

We now turn our attention to the the Newtonian limit for the non-minimally coupled fluid equations \eqref{eq:fleom}. We can in this case safely neglect the terms induced by the NMC as they would generically lead to subleading corrections. Hence, we recover the equations of motion for a minimally coupled RBEC, namely
\be\label{eq:fluidNRlimit}
\nabla_\mu(\rhom u^\mu)=0\,,\quad
  -u^\mu u_\mu=c^2+ 2U'-\frac{\hbar^2}{m^2}\frac{\Box \sqrt{\rhom}}{\sqrt{\rhom}} \md
\ee
The Newtonian limit of the continuity equation is easily obtained and gives the standard equation
\begin{equation}\label{}
\pardern{}{\rhom}{t}+\vec{\nabla}\cdot(\rhom\vec{v})=0\,.
\end{equation}
In this regime anti-bosons can be neglected \cite{p:fagnocchi} so that now the density $\rho$ is really the matter density of the non-relativistic BEC fluid. The second equation in \eqref{eq:fluidNRlimit} is an equation for the norm of $u^\mu$, not an evolution equation for the 4-vector $u^\mu$. However, as in the flat space case, we can make it such by taking a gradient on both sides of the equation. Hence we have
\begin{equation}\label{eq:fldyn}
-u^\mu \nabla_\mu u _\nu=\nabla_\nu\left(U'-\frac{\hbar^2}{2m^2}\frac{\Box\sqrt{\rhom}}{\sqrt{\rhom}}\right)\md
\end{equation}
Expanding the covariant derivative and accounting for all the terms with the same order in $1/c^2$ we obtain 
\begin{equation}\label{}
m\pardern{}{}{t}\vec{v}=- m(\vec v \cdot \vec \nabla)\vec v-\frac{1}{\rhom}\vec \nabla P(\rhom) -m\vec \nabla \Phi_N+\frac{\hbar^2}{2m}\vec \nabla\left(\frac{\nabla^2\sqrt{\rho_m}}{\sqrt{\rho_m}}\right)
\end{equation}
where we have defined $P=(U'(\rhom)\rhom-U(\rhom))$. Formally, this is exactly the dynamical fluid equation derived from the time-dependent Gross--Pitaevskii equation (see for example \cite{b:ps} for a review) when the external potential is given by the gravitational one and when the self-interaction potential is $U(\rhom)=\lambda \rhom^2$.
However, the Poisson equation \eqref{eq:poisson} is modified by the presence of the NMC and hence the gravitational potential entering in this equation is no more the standard Newtonian potential, but rather it will show a different behavior. In particular, this can be seen as an extra force which reduces/enhances the effects of gravity. 
As a concluding remark we rewrite the velocity equation as
\begin{equation}\label{velNRhealing}
m\pardern{}{}{t}\vec{v}=- m(\vec v \cdot \vec \nabla)\vec v -m\vec \nabla \Phi_N+\vec \nabla\left[\bar U'(\rhom)\left(1-\xi^2\frac{\nabla^2\sqrt{\rhom}}{\sqrt{\rhom}}\right)\right]\,,
\end{equation}
where $\bar U\equiv mU$ and $\xi^2=\hbar^2/(2m\bar U')$ is again the healing length. This form for the equation is convenient when investigating the different weights of the two pressure effects.

\section{Phenomenology}
\label{pheno}

We have discussed how the existence of a characteristic length for the system makes the BEC fluid sensible to second derivatives of the metric, thus activating the NMC. In order for this to happen, the scalar field must have undergone the phase transition \ie its temperature must be below the critical one. However, this is not enough. In fact, in order to make the mechanism effective, the characteristic scales of the BEC and of curvature are required to be of comparable size. This fact naturally splits the evolution of the scalar field into two stages which we now describe. 

\begin{figure}[t]
  \centering
\includegraphics[width=.8\linewidth]{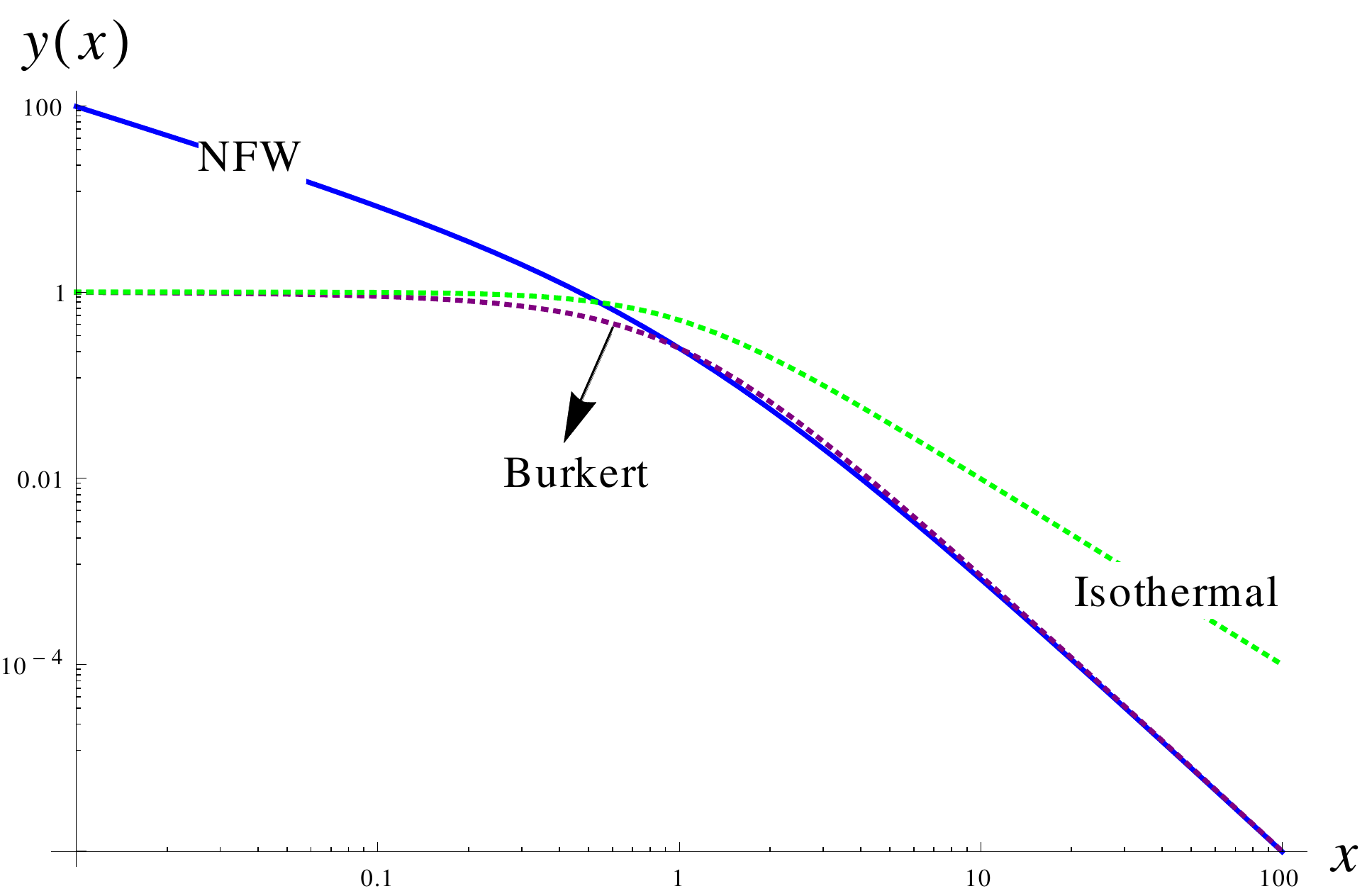} 
  \caption{Different profiles for the DM halo shape function. It is evident the cuspy behaviour of NFW profile, while both Isothermal and Burkert profiles show a constant central density profile.}
    \label{fig:Profiles_NFW_ISO_BURK}
\end{figure}

\begin{description}
\item[High temperature - low curvature phase.] Initially the scalar field is at $T>T_c$ and hence its behavior is described by a standard minimally coupled self-interacting field. In order to reproduce the observed cosmological evolution the scalar field must have a small self-interaction, \ie it must be almost pressureless, and it must be non relativistic, meaning the velocity of particles must be small compared to the speed of light. Hence, its critical temperature is given by
\be
k_BT_c^{\text{NR}}=\frac{2\pi \hbar^2}{m}\left[\frac{n}{\zeta(3/2)}\right]^{2/3}\,.
\ee
and we can neglect the anti-bosons as their number is very small in this regime \cite{p:fagnocchi}. Notice that this temperature has  the same scaling behavior as that of a pressureless fluid \cite{Takeshi:2009cy}. Hence, given that the scalar field is evolving almost as CDM, if at the onset of the epoch during which it dominates the energy budget of the universe the scalar field is above the critical temperature it is expected to be always so.  Hence, the NMC terms are not present at the background and linear perturbation level so that our model is expected to produce the same evolution given by the $\Lambda$CDM model at large scales provided that the self-interaction is small enough. 

\item[Low temperature - high curvature phase.] According to the discussion of the previous paragraph, if the boson field is above the critical temperature while it is dominating the energy density of the Universe, it will never undergo the transition. It then seems to be impossible for it to activate the non-minimal coupling. However, when we consider collapsed structures we have that the BEC is confined in an external potential provided by the gravitational one thus having a situation analogous to the one realized for a trapped BEC. In this case the critical temperature has a very different behavior becoming dependent on the characteristics of the trapping potential. For example for a BEC in an harmonic trap we have \cite{Albus:2003if}
\be
k_BT_c=\hbar \omega \left(\frac{N_B}{\zeta(3)}\right)^{1/3}
\ee
showing that now the critical temperature is proportional to the frequency $\omega$ of the trapping potential. Hence, it might be that during the collapse the critical temperature becomes high enough, inside the potential wells provided by the gravitational potential, to allow for the condensation to happen. Interestingly, an attractive interaction further increases the critical temperature, thus leading to a second welcome feature \cite{Albus:2003if}. It has to be noticed however, that such attractive interaction will generically lead to a BEC instability whenever not very small \cite{p:bosenova}. 

When the DM temperature drops below $T_c$, the scalar field condenses and develops a coherence length thus activating the NMC so that its dynamics inside the galaxy halo is given by the model described in the previous sections, provided that we identify $L^2 \propto \xi^2$, which is justified by the fact that the healing length is the only new length scale that has appeared in the system after the condensation. We will see in the next paragraph that in order to be effective in addressing the CDM issues discussed in the introduction, we need the healing length to be of the order of the kpc. Of course, a healing length of astrophysical scales might seem quite an extreme requirement (common anyway to all the DM as BEC proposals) but one has to keep in mind that DM does not need, and most probably cannot have, the same typical scales for mass, scattering length and densities, as those encountered in Earth based BEC experiments. Indeed, very weak self-interactions and very low densities might be be expected and consequently very long healing lengths are not unconceivable.

Furthermore long living quantum coherence could be assured by the very nature of DM, which feels any other matter field only gravitationally.

\item[NMC Phenomenology]

As we have described in the previous paragraph the model proposed mimics \lcdm model on large, cosmological scales while may present departures from the standard paradigm of structure formation at small scales. To sketch the possible virtues of our NMC model for DM at galactic scale we will here discuss its effect on the density profile around a spherically symmetric DM halo.

Initially the NMC is not present as the gravitational well generated by the DM distribution is not deep enough to trigger condensation. Hence, we expect that the matter infall goes on as in the standard DM scenario. N-body simulations give a density profile that is well fitted by the Navarro--Frenk--White (NFW) profile $\rho_\text{NFW}=\rho_s y_\text{NFW}(x)$ \cite{Navarro:1996gj}, where the NFW shape function $y_\text{NFW}$ is given by
\be
y_{\text{NFW}}(x)= \frac{1}{x(1+x)^2}\,,
\label{NFW}
\ee
where $x=r/r_s$, being $\rho_s$ and $r_s$ free parameters to be fitted with simulations. In the limit of small radii the NFW profile goes like $\sim x^{-1}$ which has been at the origin of much debate since in general observations are more in agreement with cored density profiles like the isothermal \cite{ISO} or Burkert \cite{Bur}. In figure \ref{fig:Profiles_NFW_ISO_BURK} are plotted the shape functions for these cases. It is evident the cuspy behaviour of the NFW function as compared with the cored one of the isothermal and Burkert functions.

At a certain point the gravitational potential may overcome the critical value for the condensation thus activating the NMC. Hence, the gravitational potential receives extra contributions through the Poisson equation \eqref{eq:poisson}. In particular, we have that for the previous choice of the density profile we can construct an effective shape function
\be
y_\text{eff}(x) =\frac{1}{x(1+x)^2}\left[1-6\epsilon \tilde L^2\frac{1}{(1+x)^2}\right]\,,
\label{shapefuncNMC}
\ee
where $\tilde L = L/r_s$. 
Interestingly, in the small radii limit
\be
\lim_{x\rightarrow 0} \rho_\text{eff} \sim  \frac{1}{x}\left[1-6\epsilon\tilde L^2 (1-2x)\right]\,,
\ee
where it can be seen that the small scale $x^{-1}$ trend is now corrected by a constant value. 

The requirement of flattening the shape function close to the center of the halo forces the sign of the NMC to take the value $\epsilon=+1$. In figure \ref{fig:Profiles_various_NMC} are reported the shape functions when the NMC is included for various values of $\tilde L$ and $\epsilon=1$, for an initial NFW profile. 
As can be easily seen for the value $\tilde L=1\sqrt{6}$ the shape function reaches an almost constant profile in the inner part of the DM halo, thus potentially relaxing some of the well known CDM issues. In physical units this correpsonds to a characteristic scale $L = r_s/\sqrt{6} \approx 0.4 r_s$. In order to get an estimate on the size of the required healing length we recall that $r_s$ is correlated to the virial mass of the halo via \cite{Salucci:2011ez}
\be
r_s \approx 8.8 \left(\frac{M_\text{vir}}{10^{11}M_\odot}\right)^{0.46} \text{kpc}\,,
\ee
thus leading to a kpc size healing length (see also \cite{Graham:2005xx}). In figure \ref{fig:Profiles_NFW_BURK_NMC} we have plotted the effective density profile for two choices of the parameter $\tilde L$ and compared them with the NFW and Burkert profile. It is clear that for the critical value the original NFW profile becomes a cored one as a consequence of the NMC corrections.
Of course, this result might seem to point toward the need of some sort of fine tuning for the model to avoid the cusp problem. It is however clear that such conclusion would be premature as in reality the healing length and the DM profile would not be independent. Only a self-consistent analysis of the developed profile for the DM condensate would provide a definitive answer in this sense. However, we feel that this preliminary analysis is enough promising to stimulate further attention in this sense.

\end{description}

\begin{figure}[t]
  \centering
\includegraphics[width=.8\linewidth]{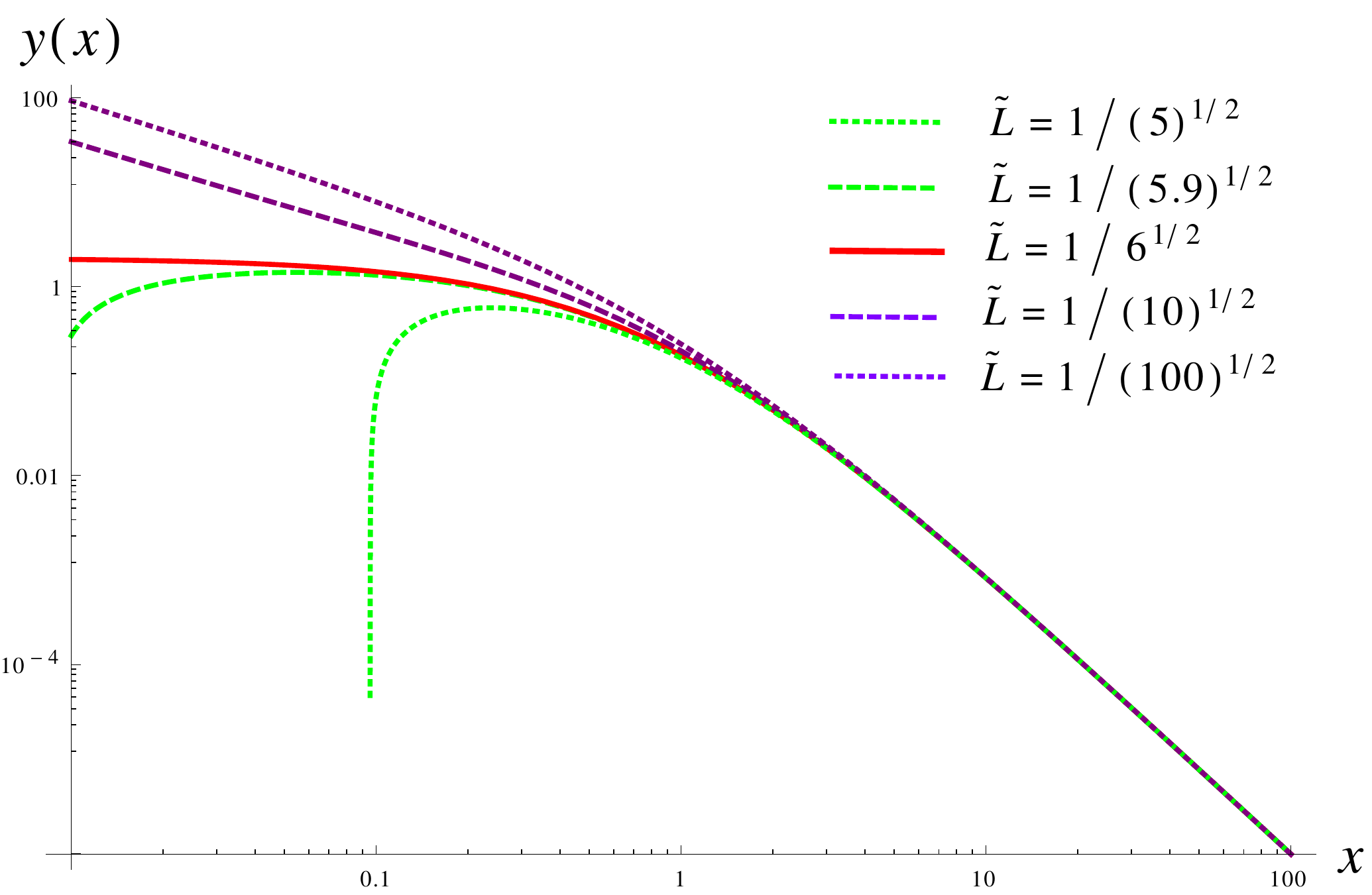} 
  \caption{Effective shape function for different values of the parameter $\tilde L$ with an initial NFW \eqref{NFW} profile. The two top lines represents small values for $\tilde L$ and results in profile closer to the initial NFW. The thick central line corresponds to the choice $\tilde L= 1/\sqrt{6}$ and shows a cored effective shape function. The two bottom lines corresponds to large values of $\tilde{L}$.}
    \label{fig:Profiles_various_NMC}
\end{figure}

\begin{figure}[t]
  \centering
\includegraphics[width=.8\linewidth]{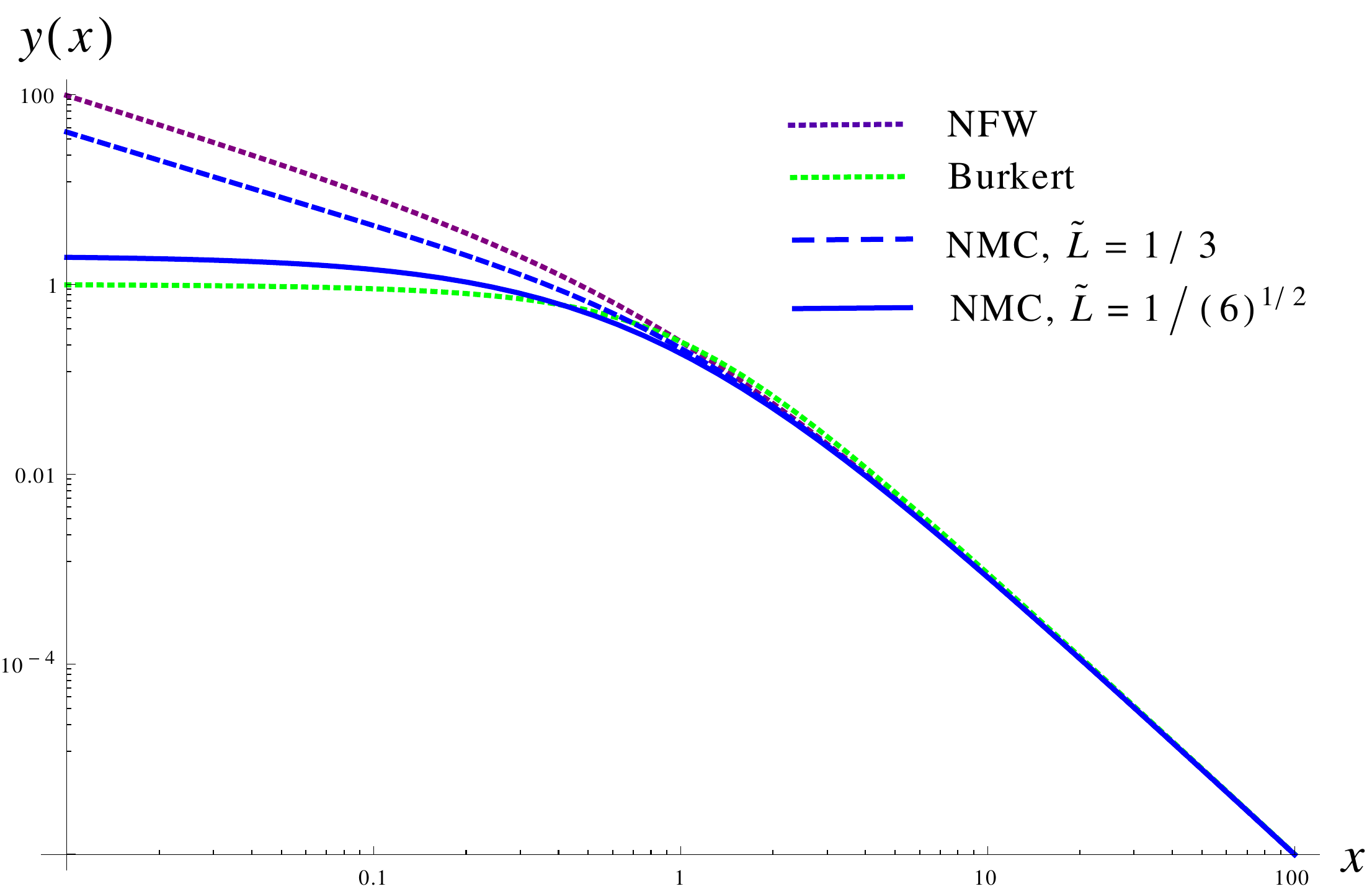} 
  \caption{NMC profiles obtained from an initial NFW profile compared with standard NFW and Burkert profile. Notice that a cored distribution is obtained for the critical value $L = r_s/\sqrt{6}$ (solid blue curve).}
    \label{fig:Profiles_NFW_BURK_NMC}
\end{figure}
To summarize, the model proposed produces a cosmology in which the dynamics of DM is given by a standard scalar field, possibly with a small attractive self-interaction which mimics standard CDM evolution until the gravitational potential is high enough to trigger the phase transition thus activating the NMC. This in turns would produce a relevant change in the dynamics in comparison with the one predicted by the $\Lambda$CDM model.

In order to be fully viable the model must be compatible with solar system and local constraints on gravitational interactions. We argue here that there are at least two reasons why we expect it to be so. On one hand, local dynamics is dominated by baryons. On the other hand, the scalar field density has to be almost constant at these scales in order to produce a cored DM density profile, thus making the NMC contributions negligible.

\section{Conclusions}
\label{conclusions}

DM still represents a daunting enigma, both for particle physics as well as for astrophysics and cosmology. Nowadays, we are assisting, under the growing pressure of the observational evidence, to a slow regress from the standard cold dark matter scenario towards warm dark matter ones or even a resurgence of many modified gravity proposals. None of the models is fully satisfactory, warm dark matter seems constrained so much to be too close to CDM for solving the aforementioned problems \cite{p:wdm}, while modified gravity models seems successful only at galactic scales \cite{p:nomond}. 

This work further explores a scenario possibly able to conjugate the successes of CDM to some ideas advanced in the modified gravity field~\cite{p:blsf,p:bls,p:bplb,p:bl}. In this scenario an almost cold DM (apart from a weak self-coupling) undergoes a dramatic phase transition in denser regions at late times forming large relativistic BECs which, having a natural coherence length --- the healing one --- develop a non-minimal coupling which in turn allows for a geometric-like influence of dark matter on the baryonic fields \cite{p:blsf}. 

To this end we have investigated the dynamics of a fully relativistic BEC with a non-minimal coupling to gravity characterised by a length scale (and of course our results can be easily reverted to the minimally coupled case by sending this scale to zero).  Within this fully relativistic framework we have looked at the Newtonian limit and found a quite simple modified Poisson equation \eqref{eq:poisson}. This is one of the most relevant results of our investigation as it shows how the non-minimal coupling associated to the presence of the healing length of the condensate modifies the non-relativistic dynamics by introducing a non-local term $L^2 \nabla^2\rhom$ on the right hand side of the equation for the gravitational potential. This term represents a departure from the standard dynamics of CDM and a possible signature of the proposed scenario.
As a first application we have shown how an initial NFW density profile is indeed flattened at small radii by the presence of the NMC as can be seen from the small radii limit of equation \eqref{shapefuncNMC} and figure \ref{fig:Profiles_NFW_BURK_NMC} thus testifying how this model may relax some of the issues of the $\Lambda$CDM at small scales.

Given the obtained results it is now possible to proceed to the next step which will consist in confronting the model with observations. There are relatively few free parameters in our theory (the mass of the DM candidate, its scattering length/self interaction strength and its average densities) and it would be interesting if reasonable values for these parameters would already allow the model to fit the wealth of observations at our disposal. For example, one might investigate what kind of constraints on the relative self-interaction strength can be derived by objects like the Musket cluster \cite{p:musk} and see if this is compatible with the healing length values required in order to fit the core cusped profiles observed in galactic holos. 
We hope that this and further analysis we be accomplished in a near future on the basis of this work.

Finally let us stress that, while our work motivations where firmly rooted in the current DM puzzles, the relevance of the model proposed goes well beyond this field. Indeed, the possible role of relativistic BEC for experimental tests of the effects of gravity on quantum physics was recently highlighted (see \eg \cite{p:fuentes}). In this direction our investigation might be used as a starting analysis of systems characterised by a general BEC-gravity coupling.

\acknowledgments
The authors wish to thank Andrea Trombettoni for illuminating discussions.
MC would like to acknowledge the support of both SISSA and the University of Trento during this research; he also wishes to thank professor Sergio Zerbini at the University of Trento for support and encouragement.

\appendix

\section{Fluid representation}
\label{Flrepr}
In this appendix we provide the relevant relations between BEC and fluid variables in the relativistic regime. In fact, if in the non relativistic regime there exists a direct relation between the two fluid representations, this is somewhat lost in the relativistic generalization. This is indeed a relevant point which deserves attention in order to avoid to misinterpret the meaning of the dynamical variables.  

Consider a BEC in the Madelung representation as discussed in section \ref{FlatRBEC}. The equation of motion and the SET are
\be
\nabla_{\mu}(\rhom u^{\mu})=0,\qquad -u_{\mu}u^{\mu}=c^{2}+2\,U'(\rhom)-\frac{\hbar^{2}}{m^{2}}\frac{\Box \sqrt{\rhom}}{\sqrt{\rhom}}
\label{app:relfluideqns}
\ee
\be
T_{\mu\nu}=\rhom u_{\mu}u_{\nu}-g_{\mu\nu}\left(\rhom\frac{c^{2}+u^{2}}{2}+\frac{\hbar^{2}}{2m^{2}}\nabla^{\alpha}\sqrt{\rhom}\,\nabla_{\alpha}\sqrt{\rhom}+U(\rhom)\right)+\frac{\hbar^{2}}{m^{2}}\nabla_{\mu}\sqrt{\rhom}\,\nabla_{\nu}\sqrt{\rhom}
\ee
where we have defined $u^{\mu}=\hbar/m\nabla^{\mu}\theta$ and $2m^{2}\rho/\hbar^{2}=\rhom$. Recall that for a complex scalar field we have the conserved current
\be
j^{\mu}=\rhom u^{\mu}
\ee
and in fact the equations of motion follows from the conservation of the current and SET. 

The above description of the BEC dynamics is referred to as the hydrodynamic version of the non-linear Klein--Gordon equation. However, the identification of these variables with standard fluid ones is not straightforward. It is clear from the second equation that the 4--vector $u^\mu$ is not the fluid 4-velocity as its norm is not unity. Moreover, the conservation equation refers to the conservation of the charge, not to the energy density. However, the fact that in the non-relativistic limit the above system resembles very closely the one for a fluid, makes the investigation of the relationships between BEC hydrodynamic variables and fluid ones worth to be explored. 

A relativistic fluid can always be described by the following SET
\be
T_{\mu\nu}^{\mathrm{fluid}}=\varepsilon v_{\mu}v_{\nu}+(\pi_{\mu}v_{\nu}+\pi_{\nu}v_{\mu})+\Psi_{\mu\nu}+\frac{1}{3}h_{\mu\nu}\mathcal{T}\,,
\ee
where $\varepsilon$ is the fluid energy density, $\pi_{\mu}$ is the 4-momentum, $\Psi_{\mu\nu}$ is a traceless tensor and $v^\mu$ is the fluid 4-velocity, normalized such that $v_{\mu}v^{\mu}=-1$. Also we can express the conserved current as
\be
j^{\mu}=n_{s}v^{\mu}\,,
\ee
where $n_{s}$ is the hydrodynamic variable associated to the charge conservation, different from $\rhom$.
By cmparing the EMT for the BEC and that for the fluid we obtain the following identifications:
\be
v_{\mu}=\frac{u_{\mu}}{\sqrt{-u^{2}}}\,,\qquad \varepsilon=-u^{2}\rhom\,,
\ee
\be
\pi_{\mu}=\frac{\hbar^{2}}{m^{2}}\left(\nabla_{\mu}\sqrt{\rhom}+v^{\alpha}\nabla_{\alpha}\sqrt{\rhom}v_{\mu}\right)v^{\beta}\nabla_{\beta}\sqrt{\rhom}\,,
\ee
\be
\mathcal{T}= -3\left(\rhom\frac{c^{2}+u^{2}}{2}+U(\rhom)\right)-\frac{\hbar^{2}}{m^{2}}\left(\nabla^{\mu}\sqrt{\rhom}\,\nabla_{\mu}\sqrt{\rhom}-2v^{\mu}v^{\nu}\nabla_{\mu}\sqrt{\rhom}\,\nabla_{\nu}\sqrt{\rhom}\right)\,,
\ee
\begin{multline}
\Psi_{\mu\nu}=\frac{\hbar^{2}}{m^{2}}\left(\nabla_{\mu}\sqrt{\rhom}\,\nabla_{\nu}\sqrt{\rhom}+(v_{\mu}\nabla_{\nu}\sqrt{\rhom}+v_{\nu}\nabla_{\mu}\sqrt{\rhom})v^{\alpha}\nabla_{\alpha}\sqrt{\rhom}+\right.\\v_{\mu}v_{\nu}v^{\alpha}v^{\beta}\nabla_{\alpha}\sqrt{\rhom}\,\nabla_{\beta}\sqrt{\rhom}
\left.-\frac{1}{3}h_{\mu\nu}\left(\nabla^{\alpha}\sqrt{\rhom}\,\nabla_{\alpha}\sqrt{\rhom}+v^{\alpha}v^{\beta}\nabla_{\alpha}\sqrt{\rhom}\,\nabla_{\beta}\sqrt{\rhom}\right)\right)\,.
\end{multline}
In particular, by substituting equation \eqref{app:relfluideqns} into $\mathcal{T}$ we get
\be
\mathcal{T}=3 p-\frac{\hbar^2}{m^2}\left[\frac{3}{2}\sqrt{\rhom}\,\Box\sqrt{\rhom}+\nabla^{\mu}\sqrt{\rhom}\,\nabla_{\mu}\sqrt{\rhom}+2v^{\mu}v^{\nu}\nabla_{\mu}\sqrt{\rhom}\,\nabla_{\nu}\sqrt{\rhom}\right]\,,
\ee
where $p=\rhom U'(\rhom)+U(\rhom)$.

Finally, one can identify
\be
n_{s}=\frac{\rho}{\sqrt{-u^{2}}}\,.
\ee

Form these results it is clear that the BEC is not a perfect fluid unless the quantum terms are subdominant. In fact in the rest frame of the BEC ($v^{\mu}=(1,0,0,0)$) we have that
\be
\pi_{\mu}=0\,, \qquad \Psi_{ij}=\frac{\hbar^{2}}{m^{2}}\left[\nabla_{i}\sqrt{\rhom}\,\nabla_{j}\sqrt{\rhom}-\frac{1}{3}g_{ij}\nabla^{k}\sqrt{\rhom}\,\nabla_{k}\sqrt{\rhom}\right]\,.
\ee
As expected in this frame there is no momentum flow but stresses are present.

If we neglect the quantum effects and use the relations between BEC and fluid variables then the SET for the BEC takes the perfect fluid form
\be 
T_{\mu\nu}=(\varepsilon + p)v_\mu v_\nu + pg_{\mu\nu}\,.
\ee

As a final remark we notice that in the non-relativistic limit fluid and BEC variables coincides, in particular the spatial part of $u^\mu$ is the real velocity of the BEC flow.

\end{document}